\pdfoutput=1
\documentclass[sigconf]{acmart}
\usepackage{booktabs}
\usepackage{multirow}
\usepackage{arydshln}
\usepackage{enumitem}
\usepackage{pifont}
\usepackage{marvosym}

\newcommand{\tabincell}[2]{\begin{tabular}{@{}#1@{}}#2\end{tabular}}

\AtBeginDocument{%
  }

\setcopyright{acmlicensed}
\copyrightyear{2024}
\acmYear{2024}
\acmDOI{10.1145/3639478.3647633}

\acmConference[ICSE'46]{2024 IEEE/ACM 46th International Conference on Software Engineering}{April 2024}{Lisbon, POR}%

\begin{document}

\title{When Large Language Models Confront Repository-Level Automatic Program Repair: How Well They Done?}

\author{Yuxiao Chen}
\affiliation{%
  \institution{Institute of Software, Chinese Academy of Sciences, China}
  \institution{University of Chinese Academy of Sciences, China}}
\email{chenyuxiao2021@iscas.ac.cn}

\author{Jingzheng Wu$\,\textsuperscript{\Letter}$}
\thanks{$\textsuperscript{\Letter}\,$Jingzheng Wu and Xiang Ling are the corresponding authors.}
\affiliation{%
  \institution{Institute of Software, Chinese Academy of Sciences, China}}
\email{jingzheng08@iscas.ac.cn}

\author{Xiang Ling$\,\textsuperscript{\Letter}$}
\affiliation{%
  \institution{Institute of Software, Chinese Academy of Sciences, China}}
\email{lingxiang@iscas.ac.cn}

\author{Changjiang Li}
\affiliation{%
  \institution{Stony Brook University, USA}}
\email{meet.cjli@gmail.com}

\author{Zhiqing Rui}
\affiliation{%
  \institution{Institute of Software, Chinese Academy of Sciences, China}
  \institution{University of Chinese Academy of Sciences, China}}
\email{zhiqing@iscas.ac.cn}

\author{Tianyue Luo}
\affiliation{%
  \institution{Institute of Software, Chinese Academy of Sciences, China}}
\email{tianyue@iscas.ac.cn}

\author{Yanjun Wu}
\affiliation{%
  \institution{Institute of Software, Chinese Academy of Sciences, China}}
\email{yanjun@iscas.ac.cn}
\renewcommand{\shortauthors}{Yuxiao et al.}

\begin{abstract}
  In recent years, large language models (LLMs) have demonstrated substantial potential in addressing automatic program repair (APR) tasks. However, the current evaluation of these models for APR tasks focuses solely on the limited context of the single function or file where the bug is located, overlooking the valuable information in the repository-level context. This paper investigates the performance of popular LLMs in handling repository-level repair tasks. We introduce RepoBugs, a new benchmark comprising 124 typical repository-level bugs from open-source repositories. Preliminary experiments using GPT3.5 based on the function where the error is located, reveal that the repair rate on RepoBugs is only 22.58\%, significantly diverging from the performance of GPT3.5 on function-level bugs in related studies. This underscores the importance of providing repository-level context when addressing bugs at this level. However, the repository-level context offered by the preliminary method often proves redundant and imprecise and easily exceeds the prompt length limit of LLMs. To solve the problem, we propose a simple and universal repository-level context extraction method (RLCE) designed to provide more precise context for repository-level code repair tasks. Evaluations of three mainstream LLMs show that RLCE significantly enhances the ability to repair repository-level bugs. The improvement reaches a maximum of 160\% compared to the preliminary method. Additionally, we conduct a comprehensive analysis of the effectiveness and limitations of RLCE, along with the capacity of LLMs to address repository-level bugs, offering valuable insights for future research.
  
\end{abstract}

\maketitle 

\section{Introduction}
Automatic program repair (APR) is an important challenge in software engineering, helping programmers significantly reduce debugging costs. Many researchers have explored various methods for APR, including pattern-based methods like TBar~\cite{liu2019tbar}, SketchFix~\cite{hua2018sketchfix} and ErrDoc~\cite{tian2017automatically}, and deep learning-based methods like CoCoNuT~\cite{lutellier2020coconut} and CURE~\cite{jiang2021cure}. Recently, the excellent generation ability of LLMs has brought new potential solutions for APR tasks. Many related studies have shown that LLMs are highly competitive in processing APR tasks~\cite{sobania2023analysis, prenner2021automatic, cao2023study, jiang2023impact}, even surpassing previously optimal methods.

However, the current evaluation of APR tasks using LLMs relies solely on the limited context of the single function or file where the bug is located. Bugs in programs can be categorized into two groups based on the size of the context they depend on for repair: function-level and repository-level~\cite{bairi2023codeplan}. For function-level bugs, the correct repair requires only providing the function where the error is located. However, due to widespread modular programming in software engineering~\cite{parnas1972criteria, sullivan2001}, there are often complex interactions or dependencies between multiple code files. This relationship can easily result in repository-level bugs, such as interface inconsistency, incorrect error handling, global variable abuse, race conditions~\cite{netzer1992race}, and more. Such repair tasks often require providing a broader repository context for repair tools. The performance of LLMs remains underexplored for repository-level repair tasks.

Our work in this paper aims to explore the performance of current popular LLMs in addressing this issue. However, the most pressing challenge is the lack of a suitable dataset. Existing datasets are either not built at repository level, such as QuixBugs~\cite{lin2017quixbugs}, or cannot accurately restore scenarios of repository-level bugs, such as Defects4~\cite{just2014defects4j}. Furthermore, datasets created too early may pose a potential risk of data leakage if used as training data for LLMs. To address this challenge, we propose a new benchmark called RepoBugs, specifically designed for evaluating repository-level APR tasks. It is built on popular open-source repositories from GitHub and contains 124 typical repository-level bugs.

\begin{figure}[h]
  \centering
  \includegraphics[width=\linewidth]{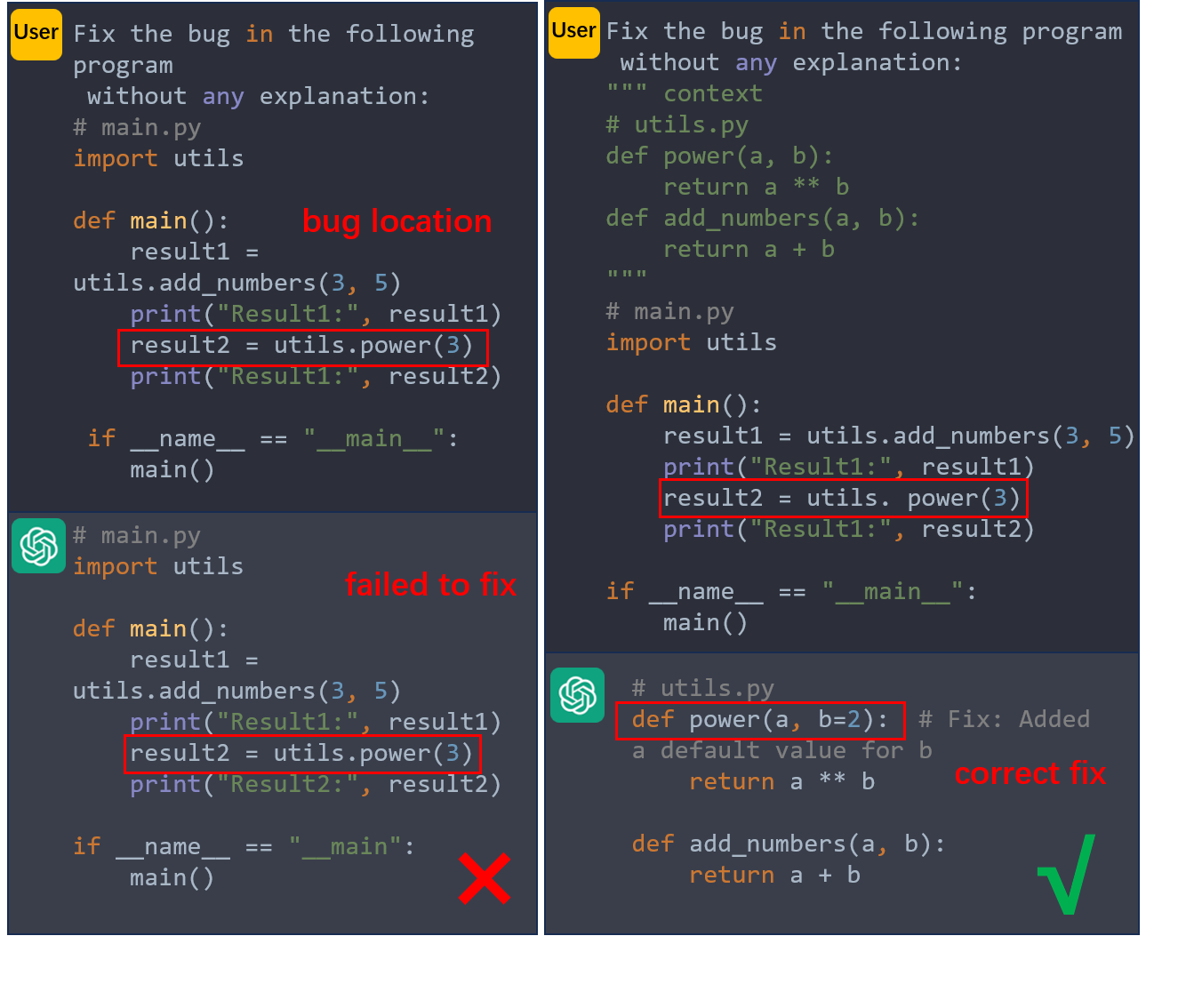}
  \caption{A simple example of using ChatGPT to handle repository-level bugs. The bug type is interface inconsistency. The number of parameter lists calling the power function is inconsistent with the function definition in \textit{utils.py}. The left figure represents the reply to ChatGPT providing the function where the bug is located as context. The right figure represents the reply after adding the repository context.}
  \label{img:example}
\end{figure}

We adopt ChatGPT from current popular LLMs for preliminary experiments. The experimental results show that if the preliminary method only use the function where the error was located as context, the repair rate of repairing on RepoBugs was only 22.58\%. This result was significantly different from the repair rate of ChatGPT on function-level bugs evaluated in other related studies~\cite{sobania2023analysis, prenner2021automatic, cao2023study, jiang2023impact}. Figure~\ref{img:example} shows a simple example. When the preliminary method only provides function-level context, ChatGPT can not perform the repair correctly. However, when we provide the complete repository as context, ChatGPT provides the correct repair result. This indicates that providing repository-level context is helpful when dealing with repository-level bugs in LLMs. Figure~\ref{img:example} illustrates a small example. When the repository size is small, we can employ a straightforward approach by considering the entire repository as the context. However, the size of the repository can be very large, and the input prompt for LLMs has an upper limit, such as a maximum token limit of 4,096 for ChatGPT. In addition, not all code in the repository is useful for the current repair task. Most code may be redundant information that interferes with the attention of the model. Therefore, it is necessary to provide precise context for the repair task of LLMs. In the field of code generation, extracting accurate context from code repositories is also an important challenge. Currently, many studies have provided solutions to this problem~\cite{lu2022reacc,zhang2023repocoder,ding2023crosscodeeval}. The methods used in these studies are roughly similar, all of which first segment the repository into slices and then obtain the most relevant fragments as context based on comparative similarity. We call this method the slice-similarity method. However, the code segments obtained through this method relying on similarity are different from the code segments that programmers refer to when making corrections to errors. Therefore, we believe that the code segments acquired through the slice-similarity method may not be well-suited for APR tasks. We also demonstrate this in subsequent experiments.

In this paper, we propose a simple and universal repository-level context extraction method (RLCE) that can extract more precise context for repository-level code repair tasks. RLCE starts with the bug location and constructs specialized prompts for handling repository-level repair tasks for LLMs by parsing repository structures, filtering code fragments, and adding auxiliary information. We evaluate three mainstream LLMs separately, and experimental results show that compared to the preliminary method, the context provided by RLCE can significantly enhance the ability of LLMs to handle repository-level bugs, with a maximum improvement of 160\%. In addition, we also conduct a comprehensive analysis of the effectiveness and limitations of RLCE, as well as the ability of LLMs to handle repository-level bugs, providing insights for future research. Our main contributions are summarized as follows:
\begin{itemize}[leftmargin=0.5cm]
\item We initially investigate the performance of popular LLMs in addressing repository-level APR tasks. The success rates of the preliminary method in repairing errors for GPT3.5 and GPT4 are only 22.58\% and 41.13\%, respectively.
\item We introduce a new benchmark, RepoBugs, which is built on popular open-source repositories from GitHub and contains 124 typical repository-level bugs. To the best of our knowledge, this is the first benchmark specifically designed for repository-level program repair.
\item The repair rate of the preliminary method is unsatisfactory, and the length of the repository often exceeds the prompt length limit of LLMs. To address these problems, we propose a simple and universal method RLCE, which provides more precise context for APR tasks and achieves over 100\% improvement in repair rates on all experimental models compared to the preliminary method.
  
\end{itemize}

\section{Benchmark Construction}
To effectively evaluate the performance of LLMs on repository-level APR tasks, the required dataset should meet the following requirements:
\begin{itemize}[leftmargin=0.5cm]
\item Each bug needs to be based on a repository context environment;
\item Bug fixing requires utilizing repository-level context;
\item The creation time of the repository is later than the collection time of current popular LLMs training data.
\end{itemize}
Moreover, conventional approaches to dataset construction in the realm of computer science, such as automatic code disruption or crawling open-source repositories, encounter challenges in accurately filtering bug types and ensuring adherence to stringent dataset quality requirements. In light of these challenges, we present a novel benchmark dataset named RepoBugs. This dataset is derived from crawled open-source Python repositories and crafted through expert manual disruption. RepoBugs comprises 124 repository-level bugs specifically addressing interface inconsistency types. Although the selection of error types is circumscribed, it is highly representative, given that interface inconsistency errors constitute the most prevalent repository-level issues~\cite{zhao2017towards}. Subsequently, we will provide a detailed exposition of the methods employed for open-source repository collection and code disruption.

\subsection{Dataset Collection}
We collect repositories from open-source projects on GitHub. To minimize the risk of leakage due to the repositories being used as training data for LLMs, we set the search date condition to later than  October 1, 2021. Additionally, we increase the number of crawled repositories. In this paper, we primarily focus on conceptual validation using the Python programming language. Consequently, we constrain the repositories considered in our work to those utilizing the Python language. Simultaneously, we ensure that each size of the repository does not exceed 1MB and that it has a minimum of 2,000 stars. We also filter out repositories with fewer than 4 Python files to adequately capture cross-file characteristics of repository-level bugs. In the end, we filter and obtain 11 repositories that meet these criteria, and detailed information is presented in Table~\ref{tab:dataset}.

\begin{table}
  \begin{tabular}{ccccc}
    \toprule
    \textbf{Index}&\textbf{Repo}&\textbf{Date}&\textbf{File}&\textbf{Sample}\\
    \midrule
    1&developer&2023/5/13&13&10\\
    2&tiktoken&2022/12/1&16&11\\
    3&gpt-migrate&2023/6/24&18&10\\
    4&starcoder&2023/4/24&7&12\\
    5&shell\_gpt&2023/1/18&16&10\\
    6&consistency\_models&2023/2/26&23&12\\
    7&musiclm-pytorch&2023/1/30&4&13\\
    8&MAE-pytorch&2021/11/1&13&12\\
    9&poe-api&2023/5/10&12&12\\
    10&ijepa&2023/6/13&15&12\\
    11&CommandlineConfig&2022/9/19&4&10\\
    \bottomrule
  \end{tabular}
  \vspace{10pt}
  \caption{Repository information in the RepoBugs dataset. \textit{Index} represents the repository number. \textit{Repo} represents the name of the repository. \textit{Date} represents the date when the repository was created. \textit{File} represents the total number of Python source files. \textit{Sample} represents the number of test samples extracted from each repository.}
  \label{tab:dataset}
  \vspace{-20pt}
\end{table}

\subsection{Dataset Generation}
The disruptive aspects of RepoBugs are manually crafted by experts with extensive programming experience. We designate the function where a call occurs as the main function and the called function as the context function, based on the characteristics of errors related to interface inconsistency. Based on existing research concerning program bugs at the repository scale~\cite{gu2019empirical,li2021large,zhang2022excepy}, we design six typical disruption rules for main and context functions, as follows:
\begin{itemize}[leftmargin=0.5cm]
\item \textbf{NRV: }Inconsistency in the number of return values between the context function and the main function. 
\item \textbf{NP: }Inconsistency in the number of input parameters between the main function and the context function.
\item \textbf{ORV: }Inconsistency in the order of return parameters between the main function and the context function.
\item \textbf{OP: }Inconsistency in the order of input parameters between the main function and the context function.
\item \textbf{CRV: }Inconsistency in the return from the context function and the requirements of the main function.
\item \textbf{CP: }Inconsistency in the input parameters between the main function and the requirements of the context function.
\end{itemize}







During the destruction process, it is important to ensure all disruptions involve interaction between two or more functions in the repository and do not introduce syntax errors that cannot pass a Python interpreter. To simplify the process, all disruptions are completed within a single line. As a result, we have obtained a total of 124 bugs in RepoBugs.

\section{Proposed Framework}

\subsection{Overall Framework}
\begin{figure*}[h]
  \centering
  \includegraphics[width=\linewidth]{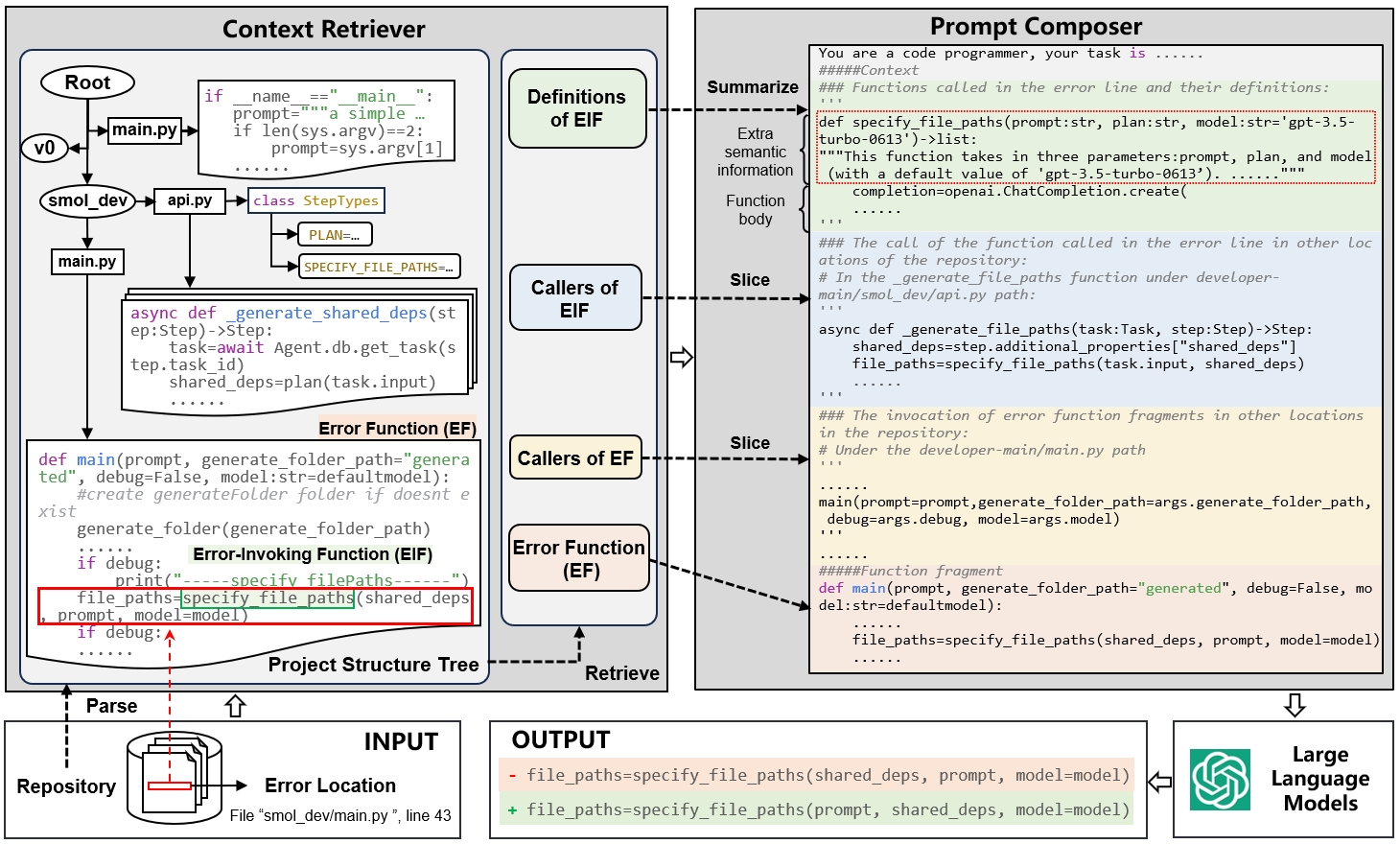}
  \caption{The overview of our repository-level context extraction method (RLCE).}
  \label{img:method}
\end{figure*}

Using LLMs to complete APR tasks at the repository level can be seen as a generation problem: the repaired code \begin{math}F^{'}\end{math} for \begin{math}F\end{math} is generated based on the function \begin{math}F\end{math} where the error is located and the context \begin{math}C\end{math} at the repository level, which can be described as \begin{math}F^{'} = M(C, F)\end{math}, where M represents the large language model used. In this step, it is crucial to obtain the repository context \begin{math}C\end{math}, and our goal is to provide more accurate context \begin{math}C\end{math} for repair tasks and generate prompts for LLMs. Figure~\ref{img:method} shows the overview of our repository-level context extraction method (RLCE). Specifically, we first use a context retriever to retrieve repository code fragments related to repair tasks (Section 3.2). Subsequently, a comprehensive large language model prompt is generated by incorporating additional information, such as appended summaries and slices (Section 3.3).

\subsection{Context Retriever}
The core of RLCE lies in addressing the questions of where to retrieve from the repository and what kind of context to obtain. To achieve this objective, we design and implement a context retriever, which is a static code analysis tool capable of automatically parsing the repository into code segments based on its structure. It retrieves segments relevant to the repair task based on error localization information. Given that repository projects often have complex structures due to dependencies among files, our tool needs to possess the capability to analyze the structure of the repository for a clearer understanding of the relationships between its components. The overall structure of the context retriever is illustrated in Figure~\ref{img:method}, consisting primarily of two key steps: (1) parsing repository files and constructing the project structure tree and (2) conducting retrieval in the project structure tree based on error location to obtain the required code segments.

\textbf{Build project structure tree: } During the construction of the project structure tree, our primary focus revolves around five types of entity nodes, namely: directories, files, classes, functions, and global variables. The connections between these nodes adhere to the original structural relationships within the repository. For example, the partial project structure tree of the \textit{developer} repository is illustrated in Figure~\ref{img:method}. Specifically, a project structure tree originates from a root node, with its child nodes encompassing subdirectories and files under the root directory of the repository. The child nodes of file entities include globally defined variables, classes, and functions. The leaf nodes of the project structure tree are restricted to function nodes or variable nodes, encompassing the code where functions or variables are defined. In addition to structural information, for the sake of facilitating subsequent retrieval processes, if a file calls functions, variables, or classes defined in another file, markers need to be placed on the file node for reference.

\textbf{Retrieve code segments: }As illustrated in Figure~\ref{img:method}, the errors for the targeted code fix task are localized within one or several lines of code, referred to as the ``error location''. Before retrieval, the context retriever tool needs to analyze and extract the functions and global variables called within the error location, which we term Error-Invoking Functions (EIF). Subsequently, we define four types of context sources to determine where the retriever should extract code segments from the project structure tree as part of the context:
  \begin{itemize}[leftmargin=0.5cm]
    \item \textbf{Definitions of EIF: }Retrieve code segments containing the definitions of the extracted Error-Invoking Functions within the repository scope.
    \item \textbf{Callers of EIF: }Search other occurrences of the Error-Invoking Function within the repository (excluding the error location) to obtain code segments containing their calling locations.
    \item \textbf{Error Function (EF): }The function containing the error location. 
    \item \textbf{Callers of EF: }Examine if the Error Function is called elsewhere in the repository, and if so, retrieve code segments containing the calling locations.
  \end{itemize}

  A null value will be returned During the retrieval process if a particular context source is absent. Formally, the context retriever constructs a project structure tree \begin{math}T\end{math} for the repository and gathers a collection of code segments \begin{math}C_{repo} = R(EL, T)\end{math}, where \begin{math}C_{repo}\end{math} encompasses all code segments from the four context sources. 

\subsection{Prompt Composer}
The primary function of the prompt composer is to further process the code segments obtained by the context retriever and merge them with templates from different prompt strategies to generate the final prompt for the large language model. This can be represented as \begin{math}C = P(C_{repo})\end{math}, where \begin{math}C\end{math} represents the repository-level context in the final prompt, and \begin{math}P\end{math} denotes the process of handling the collection of code segments. As depicted in Figure~\ref{img:method}, for each context source, our processing approach is as follows:

\textbf{Definitions of EIF: }For this part, we attach extra semantic information to EIF to enhance the model's understanding of the function purposes and parameter meanings. Semantic information comprises two components: function signature and function summary. The function signature includes the function name, the type of each parameter in the parameter list, and the type of return value. The function summary provides an overview of the main functionality. Any model capable of generating code summaries and signatures can be used. For the sake of simplicity and leveraging the outstanding performance of LLMs in code summarization, our experiments choose the LLMs as the generation model.

\textbf{Callers of EIF: }For this part, we employ a slicing approach for processing. Since the error location contains calls to EIF, calls to EIF in other locations within the repository are likely to have valuable references for error correction. Therefore, the most useful information in code segments from Callers of EIF for the repair task is primarily concentrated around the statements where EIF are invoked. To minimize the introduction of excessive redundant information, we adopt a slicing approach, preserving the content of the statements before and after the invocation, each with a context window of five lines.

\textbf{Callers of EF: }For Callers of EF, providing useful contextual information about the Error Function may contribute to a better understanding of LLMs. Therefore, similar to Callers of EIF, the approach for this section also employs the same slicing method. 

\textbf{Error Function (EF): }In the context of Error Function, no additional processing has been applied; instead, they are incorporated directly into the prompt context.

It is worth noting that, in this paper, we primarily focus on conducting our experiments using programs written in the Python language. However, the design principles of the context retriever can be extended to other programming languages. Examples of our method of repairing bugs in the Java programming language can be found in Appendix~\ref{case:two}.

\section{Evaluation Setup}

\subsection{Models}
We select three representative models from the currently most popular LLMs for our experiments.

\textbf{GPT3.5}~\cite{ouyang2022training} is developed by OpenAI. It has strong language comprehension and generation capabilities through large-scale data training. The model we select in our experiment is GPT-3.5 Turbo, which supports a maximum token sequence length of 4K. The training data is up to September 2021. 

\textbf{PaLM2}~\cite{anil2023palm} is a new generation big language model 
  launched by Google. It has advanced reasoning ability and natural language generation ability. In our experiment, we select the text bison-001 model, which supports a maximum of 8K tokens input and 1K output, with a knowledge cutoff date of mid-2021. 

\textbf{GPT4}~\cite{openai2023gpt4} stands as one of the most powerful LLMs currently released by OpenAI. In handling intricate tasks, GPT-4 exhibits greater reliability and creativity. For our experiments, we select the gpt-4-0613 model, which supports a maximum token length of 8K and was trained with data up to September 2021. 

As these models are not open-source, we obtain responses through their APIs.

\subsection{Prompt Generation}
Due to the significant impact of prompt engineering on the performance of LLMs~\cite{liu2023pre, white2023prompt}, to fully evaluate their performance, we adopt the following three most commonly used prompt strategies in various tasks (for simplicity, specific prompt designs can be found in the Appendix~\ref{app:prompt design}), including zero-shot, one-shot, and chain of thought (CoT).

\textbf{Zero-shot}~\cite{brown2020language}: This strategy does not provide any example of the model during the inference process, only natural language instructions describing the task. This method can minimize the limited prompt length and provide more contextual capacity for repair tasks, but it also faces difficulties in understanding task formats and other issues. We use two types of instructions, \textit{Simple} and \textit{Detail}, in the experiment. The Simple format instruction describes the task in an extremely concise language, while the Detail format is more specific, requiring the large language model to assume that it is a programmer completing a task of fixing bugs from the repository using context.

\textbf{One-shot}~\cite{brown2020language}: This strategy is similar to zero-shot but allows for an example other than natural language instructions that describe the task. We use the same instruction as the Detail method in the zero-shot strategy in the experiment and add a complete repair example.

\textbf{CoT}~\cite{wei2022chain}: Previous studies have shown that the CoT strategy can significantly enhance the reasoning ability of LLMs. The process of automatically fixing bugs can be seen as an inference process. To investigate the efficacy of the CoT strategy in the field of repository-level APR, we propose a straightforward zero-shot-CoT~\cite{kojima2022large} approach in this paper. This approach decomposes the repair task into three distinct logical steps: first, identifying the root cause of errors by integrating contextual information; second, devising targeted solutions based on the identified error causes; and finally, generating the comprehensive repaired code. Our instructional prompt guides the model to systematically engage in these three steps, providing explicit error explanations, repair strategies, and the resultant fixed code, respectively.

\subsection{Compared Repair Baselines}

  \textbf{Preliminary method}: 
  A key contribution of our RLCE is the extraction of more precise repository-level context for repair tasks. To demonstrate the effectiveness of our method, we implement a preliminary method as the baseline that only provides the function itself where the error is located as the context. This simulates the existing preliminary method that leverages LLMs to address function-level APR tasks.

  \textbf{Slice-similarity method}:  To explore whether the slice-similarity method applies to the field of APR, our experiment reproduced the Retrieval Model used in the RepoCoder~\cite{zhang2023repocoder} method. Specifically, we set the slicing window to 10, use the sparse word bag model as the vector representation model, and use the Jaccard algorithm to calculate the similarity between the segment where the error is located and other segment vectors in the repository. Finally, 5 segments with the highest similarity are selected as the context.

\subsection{Evaluation Metrics}
Due to the high cost of running the repository and designing test cases, as well as the diversity of bug-fixing solutions, effective fixing accuracy cannot be achieved through precise matching (example can be found in the Appendix~\ref{case:one}). Therefore, to ensure the accuracy of the evaluation results, we ultimately adopt a manual evaluation method, and the evaluation results were provided by two experts who have more than 5 years of experience in Python programming. We will divide the evaluation indicators into four items to fully evaluate the return results of the large model. The specific evaluation criteria are as follows. When evaluating, if it meets the criteria, it is marked as 1; otherwise, it is marked as 0:
\begin{itemize}[leftmargin=0.5cm]
  \item \textbf{Related reply}: The return result is not empty and is related to the repair task in the instruction.
  \item \textbf{Correct format}: The returned result is in the expected format and the content is complete without any duplicate or redundant content.
  \item \textbf{Correct repair}: The returned result contains the correct fix for the error.
  \item \textbf{Correct explanation}: This item is specialized for the CoT prompt strategy, and the criteria are that the returned result includes a correct explanation of the cause of the error.
\end{itemize}                                                                                                                                                                                                                                                                                                            
In the specific evaluation process, the two experts first have a detailed discussion on the evaluation criteria to ensure both reach a consistent understanding. Then, the two experts independently evaluated all the experimental results. Subsequently, the evaluation results of the two individuals are compared, and any minor discrepancies were addressed through further discussion and reassessment by both experts. This iterative process ensured the attainment of a unanimous final result.
\section{Results and Analysis}
\subsection{RLCE Outperforms the Baselines}
We present the results of our experiments in Table~\ref{tab:mainexp}, where each cell represents the proportion of samples passing the corresponding evaluation metric among all samples. Our RLCE method exhibits a significant improvement compared to the other two baselines. Regarding the relevance of model responses and the correctness of format, all methods and models generally generate responses relevant to the questions and conform to the expected format. In terms of repair rate, we observe that all models perform poorly when employing the preliminary method. For instance, even the superior GPT4 achieves only a 41.43\% success rate. This indicates that even state-of-the-art LLMs struggle to accomplish repository-level repair tasks with only limited context at the function level, as they fail to provide sufficient information for repair tasks. After using the RLCE method to provide repository-level context, the repair rate improvement compared to the preliminary method has reached over 100\%, with the GPT3.5 reaching the highest of 160\%. This underscores the necessity of repository-level context for such code repair tasks. 

Additionally, across the experiments with the three models, the repair rate of the slice-similarity method does not surpass our RLCE method. For example, in the GPT4 model, the respective improvement rates compared to the preliminary method are 20\% and 100\%. This suggests that the enhancement provided by the context of this method is limited. The slice-similarity method exploits code repetition in the repository, retrieving code segments similar to the error location to aid in error repair. However, relying solely on similar code makes it challenging to reconstruct the actual execution-time context before and after the error location, leading LLMs to struggle in correctly inferring the reasons for errors.

\begin{table*}[]
  \begin{tabular}{@{}llcccccc@{}}
  \toprule
  \multirow{3}{*}{\textbf{Model}}  & \multirow{3}{*}{\textbf{Metric}} & \multicolumn{6}{c}{\textbf{Method}}                                                                                                                                          \\ \cmidrule(l){3-8} 
                                   &                                  & \multirow{2}{*}{\textbf{Preliminary}} & \multirow{2}{*}{\textbf{Slice-similarity}}       & \multicolumn{4}{c}{\textbf{RLCE}}                                                 \\ \cmidrule(l){5-8} 
                                   &                                  &                                       &                                                  & \textbf{Simple} & \textbf{Detail} & \textbf{One-shot} & \textbf{CoT} \\ \midrule
  \multirow{4}{*}{\textbf{GPT3.5}} & Related reply                   & \textbf{1}                             & \textbf{1}                                       & \textbf{1}        & \textbf{1}               &\textbf{1}            & 0.9919                    \\
                                   & Correct format                   & \textbf{0.9597}                       & 0.9516                                           & 0.9274          & 0.9516          & \textbf{0.9597}   & \textbf{0.9597}           \\
                                   & Correct repair                   & 0.2258                                & 0.3387                                           & 0.4113          & 0.5645          & \textbf{0.5968}   & 0.5161                    \\
                                   & Correct explanation              & -                                     & -                                                & -               & -               & -                 & \textbf{0.5284}           \\ \midrule
  \multirow{4}{*}{\textbf{PaLM2}}  & Related reply                  & \textbf{0.8871}                       & 0.8468                                           & 0.8387          & 0.8629          & 0.871             & 0.879                     \\
                                   & Correct format                   & \textbf{0.8548}                       & 0.7903                                           & 0.7984          & 0.8064          & 0.8387            & 0.5242                    \\
                                   & Correct repair                   & 0.2177                                & 0.2419                                           & \textbf{0.4272} & 0.3952          & 0.4032            & 0.2742                    \\
                                   & Correct explanation              & -                                     & -                                                & -               & -               & -                 & \textbf{0.1774}           \\ \midrule
  \multirow{4}{*}{\textbf{GPT4}}   & Related reply                   & 0.9677                                & \textbf{1}                                       &0.9919           & 0.9919          & 0.9839            & \textbf{1}                \\
                                   & Correct format                   & 0.9677                                & 0.9758                                           &0.9839           & 0.9758          & 0.9839            & \textbf{1}                \\
                                   & Correct repair                   & 0.4113                                & 0.4919                                           &0.7742           & 0.7581          & \textbf{0.8145}   & 0.75                      \\
                                   & Correct explanation              & -                                     & -                                                &-                & -               & -                 & \textbf{0.7742}           \\ \bottomrule
  \end{tabular}
  \vspace{8pt}
  \caption{Evaluation results of different models and methods on RepoBugs. The values in the cells represent the proportion of samples that passed the corresponding evaluation metrics out of the total number of samples. The cell data corresponding to the best-performing method in each row is bolded. Detailed descriptions of the \textit{Preliminary} and \textit{Slice-similarity} methods can be found in Section 4.3, both employing the one-shot prompting strategy.}
  \label{tab:mainexp}
\end{table*}

\subsection{Impact of Prompt Strategies on RLCE}
It is worth noting that the performance of the RLCE method is influenced by the use of different prompt strategies. As shown in Table~\ref{tab:mainexp}, a comparison of different models reveals the most pronounced performance fluctuation in GPT3.5 due to the prompting strategy. Specifically, the one-shot strategy yield a repair rate improvement of over 37\% compared to the Simple method with the zero-shot strategy. In practical repair tasks, it is desirable for the model to exhibit low sensitivity to prompting strategies. Because, in situations with similar contexts, stable responses of the model to various prompting strategies can reduce the uncertainty of repair outcomes and mitigate the exploration costs associated with finding appropriate prompting strategies. In summary, considering both response stability and accuracy, GPT4 demonstrates optimal performance.

\subsection{Correct Explanation is Important for CoT}
\begin{figure}[h]
  \centering
  \includegraphics[width=\linewidth]{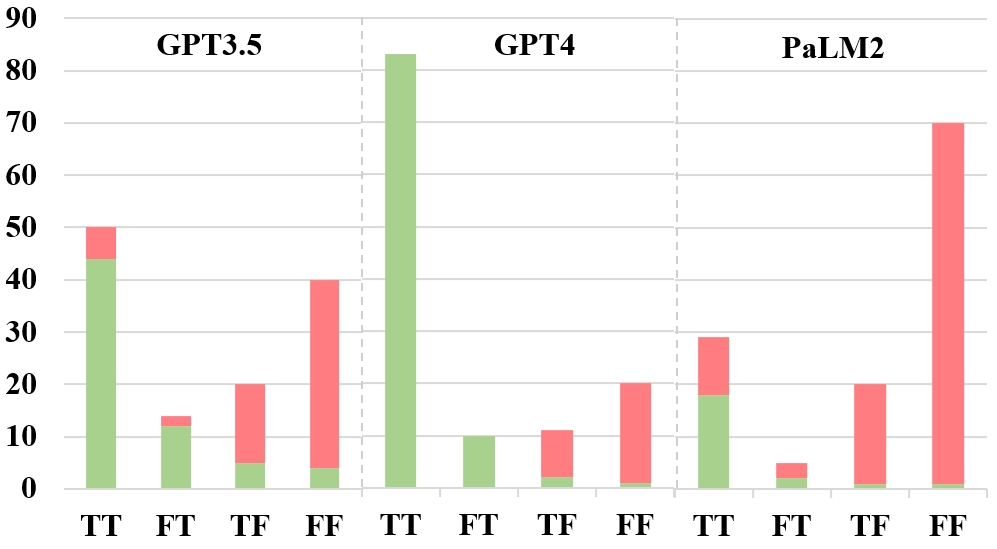}
  \caption{Statistical categorization of repair outcomes using the detail and CoT methods for three models, along with CoT-generated explanations. The horizontal axis comprises four categories, each composed of two uppercase letters, T or F, representing whether the repair results using the Detail and CoT methods are correct (e.g., FT denotes all cases where Detail repair is incorrect and CoT repair is correct). The vertical axis represents the number of cases for each category. In each category, red indicates instances where the CoT method provided incorrect explanations, while green signifies correct explanations.}
  \label{img:de-cot}
\end{figure}

Analyzing the data in Table~\ref{tab:mainexp}, we find that the repair rate of all models employing the CoT strategy did not meet our expectations, with a notable decrease in performance even on PaLM2. The GPT4, which exhibited the best performance overall, also experienced a slight decline in repair accuracy after adopting the CoT strategy. To investigate the reasons behind this, we categorize and statistically analyze the repair outcomes of the three models using both the Detail and CoT methods, along with the explanations generated by CoT. The results are presented in Figure~\ref{img:de-cot}. It is evident from the data that a significant correlation exists between the repair and explanation aspects for all models. In cases where the CoT method was successfully repaired (clusters denoted by the second uppercase letter T), the proportion of correct explanations was exceptionally high. Moreover, as the inference capabilities of the models improved, this trend became more pronounced. The statistical results for GPT4 indicate that all successfully repaired cases also had accurate explanations. When observing the three sets of TF clusters, corresponding to cases where the Detail method successfully repaired while the CoT method did not, the proportion of incorrect explanations was significantly higher than that of correct explanations. From the analysis above, we hypothesize that in the CoT method, incorrect interpretations can introduce significant interference, leading to erroneous outcomes. Using models with enhanced inference capabilities or guiding models to generate more accurate explanations may potentially improve the repair rate.
\subsection{Validity of Context Sources}
Our research results indicate that the RLCE method can significantly enhance repository-level program repair tasks. To explore the correlation between the four context sources mentioned in Section 3.2 and extra semantic information with code repair performance, we conduct a set of ablation experiments. In these experiments, we remove three context sources and extra semantic information, excluding EF due to the inclusion of error localization. The results of the experiments can be found in Table~\ref{tab:source}. The models selected for the experiments are GPT3.5 and GPT4, both showing better overall performance, with all strategies adopting the one-shot approach.

\begin{table}[htbp]
\setlength\tabcolsep{1pt}
\small
  \begin{tabular}{@{}lccccccc@{}}
    \toprule
                                      & \multicolumn{4}{c}{\textbf{Context Source}}                                                                                                   & \multicolumn{3}{c}{\textbf{Evaluation}}                                     \\ \cmidrule(l){2-8} 
    \multirow{-2}{*}{\textbf{Model}}  & \textbf{\tabincell{c}{Summa\\rize}}  & \textbf{\tabincell{c}{Callers\\of EF}}                      & \textbf{\tabincell{c}{Definitions\\of EIF}} & \textbf{\tabincell{c}{Callers\\of EIF}}                     & \textbf{\tabincell{c}{Related\\reply}} & \textbf{\tabincell{c}{Correct\\format}} & \textbf{\tabincell{c}{Correct\\repair}} \\ \midrule
                                      & {\color[HTML]{32CB00} \textbf{\checkmark}} & {\color[HTML]{32CB00} \textbf{\checkmark}} & {\color[HTML]{32CB00} \textbf{\checkmark}} & {\color[HTML]{FE0000} \textbf{×}} & 1                       & 0.9597                  & 0.5806                  \\
                                      & \textbf{-}                        & {\color[HTML]{32CB00} \textbf{\checkmark}} & {\color[HTML]{FE0000} \textbf{×}} & {\color[HTML]{32CB00} \textbf{\checkmark}} & 1                       & 0.9355                  & 0.2742                  \\
                                      & {\color[HTML]{32CB00} \textbf{\checkmark}} & {\color[HTML]{FE0000} \textbf{×}} & {\color[HTML]{32CB00} \textbf{\checkmark}} & {\color[HTML]{32CB00} \textbf{\checkmark}} & 1                       & 0.9597                  & 0.5565                  \\
                                      & {\color[HTML]{FE0000} \textbf{×}} & {\color[HTML]{32CB00} \textbf{\checkmark}} & {\color[HTML]{32CB00} \textbf{\checkmark}} & {\color[HTML]{32CB00} \textbf{\checkmark}} & 1                       & 0.9597                  & 0.5403                  \\
    \multirow{-5}{*}{\textbf{GPT3.5}} & {\color[HTML]{32CB00} \textbf{\checkmark}} & {\color[HTML]{32CB00} \textbf{\checkmark}} & {\color[HTML]{32CB00} \textbf{\checkmark}} & {\color[HTML]{32CB00} \textbf{\checkmark}} & 1                      & 0.9597                  & \textbf{0.5968}         \\ \midrule
                                      & {\color[HTML]{32CB00} \textbf{\checkmark}} & {\color[HTML]{32CB00} \textbf{\checkmark}} & {\color[HTML]{32CB00} \textbf{\checkmark}} & {\color[HTML]{FE0000} \textbf{×}} & 0.9839                  & 0.9839                  & 0.7742                  \\
                                      & \textbf{-}                        & {\color[HTML]{32CB00} \textbf{\checkmark}} & {\color[HTML]{FE0000} \textbf{×}} & {\color[HTML]{32CB00} \textbf{\checkmark}} & 0.9919                  & 0.9677                  & 0.5000                  \\
                                      & {\color[HTML]{32CB00} \textbf{\checkmark}} & {\color[HTML]{FE0000} \textbf{×}} & {\color[HTML]{32CB00} \textbf{\checkmark}} & {\color[HTML]{32CB00} \textbf{\checkmark}} & 0.9839                  & 0.9839                  & 0.7661                  \\
                                      & {\color[HTML]{FE0000} \textbf{×}} & {\color[HTML]{32CB00} \textbf{\checkmark}} & {\color[HTML]{32CB00} \textbf{\checkmark}} & {\color[HTML]{32CB00} \textbf{\checkmark}} & 0.9839                  & 0.9839                  & 0.7823                  \\
    \multirow{-5}{*}{\textbf{GPT4}}   & {\color[HTML]{32CB00} \textbf{\checkmark}} & {\color[HTML]{32CB00} \textbf{\checkmark}} & {\color[HTML]{32CB00} \textbf{\checkmark}} & {\color[HTML]{32CB00} \textbf{\checkmark}} & 0.9839                  & 0.9839                  & \textbf{0.8145}         \\ \bottomrule
    \end{tabular}

  \vspace{10pt}
  \caption{Comparison of the effects of different context sources in the Context.}
  \label{tab:source}
  \vspace{-25pt}
\end{table}

From the results in Table~\ref{tab:source}, it is evident that both models perform well in terms of response format and relevance. In terms of accuracy, the experimental groups utilizing the complete context sources achieve the highest repair accuracy. However, the repair rate of different context sources on the repair results varies. If the context source of definitions of EIF is lacking, both models exhibit a significant decrease in repair rates, with GPT3.5 even dropping by more than half. This indicates that the definitions of EIF context source provide the most important information for the repair task. Moreover, the extra semantic information does not noticeably enhance the success rate. It suggests that the useful information extracted from it might be limited for LLMs.

\subsection{Error Types Affect Repair Effectiveness}
\begin{figure*}[h]
  \centering
  \includegraphics[width=\linewidth]{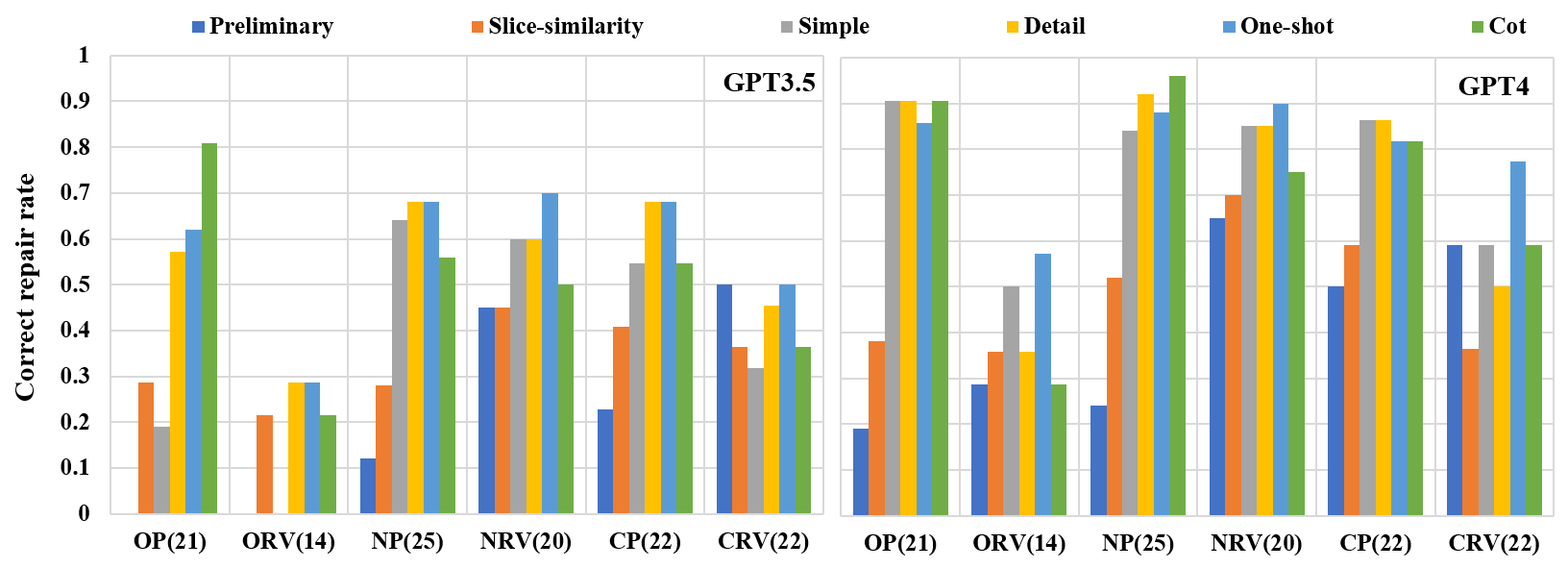}
  \caption{The results of various prompt strategies employed by GPT-3.5 and GPT-4 are statistically compiled in terms of correction accuracy based on error categories. The horizontal axis denotes the six categories of errors, with specific definitions provided in Section 2.2. The total number of samples corresponding to each category is enclosed in parentheses. Different colors of bars represent distinct methods, while the vertical axis represents the repair accuracy.}
  \label{img:error type}
\end{figure*}

In Section 2.2, we present six distinct disruption rules. Errors arising from these diverse disruption rules often necessitate varying analytical approaches of LLMs to repair. To investigate the differences in the repair rate of LLMs when confronted with different error types, we categorize the results of distinct prompt strategies based on error categories and compute the correction accuracy of GPT-3.5 and GPT-4 model, as depicted in Figure~\ref{img:error type}.

Analysis of Figure~\ref{img:error type} reveals a consistent trend in the performance of both models when addressing different error types. Both models exhibit poorer performance when faced with ORV and CRV error types. Upon examining instances of these two error types, we observe that, compared to other categories, these errors are more subtle and challenging to repair, mainly in two aspects. Firstly, ORV and CRV errors involve parameter mismatch or incorrect order between interfaces, requiring a deep understanding of the specific meaning of each parameter and the functionality of the function. This requires a deep semantic understanding. In contrast, the mismatch in the number of parameters can often be identified through simple syntax checks. Secondly, in real-world repositories, variable names in most cases align between parameters and arguments (We generally use the word `parameter' for a variable named in the parenthesized list in a function definition, and `argument' for the value used in a call of the function~\cite{1988The}), enabling the large language model to identify some obvious errors by comparing the parameter list in the function definition position with the argument list in the invocation position. In contrast, inconsistencies in the return values, as seen in ORV and CRV types, demand the large language model to comprehend the functionality of function and even the meaning of each variable during the parameter-passing process.

\subsection{Long Prompt vs. Short Prompt}
\begin{figure}[!h]
  \centering
  \includegraphics[width=\linewidth]{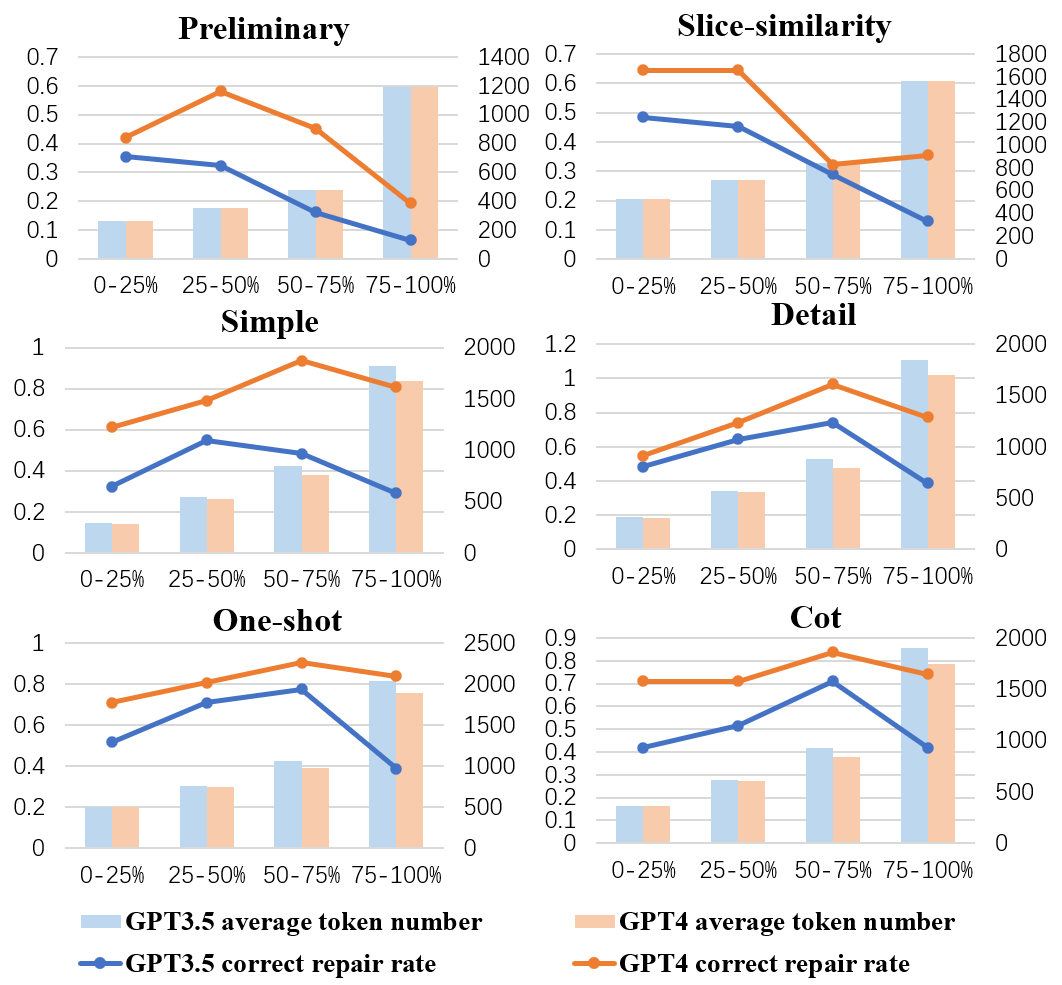}
  \caption{The relationship between different prompt lengths and repair accuracy is depicted through six subfigures. Each subfigure represents a distinct prompt strategy. All cases within each prompt strategy are evenly divided into four subsets based on prompt length (with a total dataset size of 124, resulting in 31 cases per subset). In each subfigure, bars represent the average length of tokens in the prompt, while the line graph illustrates the repair accuracy.}
  \label{img:prompt}
  \vspace{-10pt}
\end{figure}

The prompt length of LLMs is subject to an upper limit. Does this imply that one should aim to incorporate as much information from the repository as possible within the confines of the prompt length? We classify the experimental results of GPT-3.5 and GPT-4 models using the one-shot method based on the length of the prompt. The statistical results are illustrated in Figure~\ref{img:prompt}. It is observed that as the prompt length increases, both GPT3.5 and GPT4 exhibit a noticeable decline in repair accuracy. This reflects that the performance of LLMs varies when handling contexts of different scales. We posit that, for program repair tasks, a longer prompt may provide more information to the large language model, but simultaneously, it may also distract the model, increasing the difficulty of repair. Therefore, the longer length of the provided context does not lead to better results; instead, it should aim to provide precise context to enhance the density of useful information for the APR task.

\section{Threats to Validity}
\textbf{Internal Validity}: In our experiments, we select three major LLMs (GPT3.5, GPT4, and PaLM2). However, different models usually differ in various aspects such as parameter count, training data, and fine-tuning methods. These differences may impact the repair rate of our method. For instance, if a large language model is inadequately trained on a specific programming language, its performance for that language may be suboptimal. Additionally, in our experiments, we standardize the temperature parameter across all LLMs to 0, ensuring stable outputs when facing the same prompt. Increasing the temperature parameter leads to unstable model outputs, potentially influencing the final results.

\textbf{External Validity}: The most significant factor influencing external validity is the choice of datasets. Firstly, if the creation time of repositories in the dataset predates the cutoff time of the training data, there is a high probability that the repository has already been learned as part of the training data, potentially impacting the final results. Secondly, an increase in the scale and complexity of repositories may escalate the cost and difficulty of context retrieval, affecting the reliability of the results. Additionally, in our experiments, we primarily focus on the error type of interface inconsistency, as it is common at the repository level and can typify the language model's ability to handle repository-level errors. However, given that RepoBugs is the only dataset known to us specifically designed for repository-level error repair, the generalization of our approach to datasets involving other error types may vary.

\section{Related Work}
Automatic program repair (APR) is a crucial research problem in the field of software engineering, representing a significant research direction to reduce the cost of software maintenance and enhance software reliability. Since the introduction of an APR framework by Arcuri and Yao in 2008~\cite{Arcuri_Xin_Yao_2008}, the field of APR has undergone rapid development. Early approaches to APR predominantly relied on traditional methods; for instance, GenProg~\cite{Weimert_2009} employed an extended form of genetic programming to evolve program variants and validated the effectiveness of repairs through test cases. TBar~\cite{liu2019tbar} explored template-based APR methods, assessing the effectiveness of different repair patterns through experiments. LSRepair~\cite{Liu_Bissyande_2018} addressed program errors by conducting real-time searches for repair components within existing code repositories.

In recent years, a substantial amount of research has shifted towards leveraging machine learning techniques, particularly deep learning, for program repair. Researchers generate repair solutions by learning from extensive repair data in code repositories. For example, DeepFix~\cite{Gupta_Pal_Kanade_Shevade_2022} utilized a multi-layer sequence-to-sequence neural network to fix common programming errors without relying on external tools for locating or repairing. SequenceR~\cite{Chen_Kommrusch_2021} combined the encoder/decoder architecture with a copying mechanism to overcome the challenge of large vocabulary in source code. CURE~\cite{jiang2021cure} integrated pre-trained GPT models for programming languages with translation models, introducing a Context-aware Targeted Search strategy. SGEPR~\cite{xiaoyu} uses a novel intermediate representation named sequence code property graph (SCPG) to model program semantic information. Recently, the remarkable comprehension and generation capabilities of LLMs have attracted widespread attention. Nan, Liu, and others~\cite{jiang2023impact} compared ten code language models and four deep learning APR techniques across four APR benchmarks. The experimental results demonstrated the competitive repair abilities of code language models.
\section{Conclusion}
In summary, we conduct a pioneering evaluation of the capabilities of major existing LLMs in handling repository-level repair tasks. We introduce a benchmark dataset, RepoBugs, along with a straightforward and versatile repository-level context extraction method, RLCE. RLCE, leveraging repository structure parsing and relevant context retrieval, offers more precise context for repository-level repair tasks. Experiments on the RepoBugs benchmark indicate that RLCE significantly enhances the performance of LLMs in repository-level program repair tasks. Furthermore, we conduct a detailed analysis of the experimental results from aspects such as context sources, error types, and prompt length, providing valuable insights for future research. RLCE has the potential to empower LLMs to offer efficient and accurate guidance for addressing errors encountered in actual development processes.

\begin{acks}
This paper is supported by the Strategic Priority Research Program of the Chinese Academy of Sciences under Grant No. XDA0320401 and the National Natural Science Foundation of China under No. 62202457. This paper is supported by YuanTu Large Research Infrastructure.
\end{acks}

\appendix
\section{Appendix}
\subsection{Prompt design}
\label{app:prompt design}
\begin{figure}[h]
  \centering
  \includegraphics[width=\linewidth]{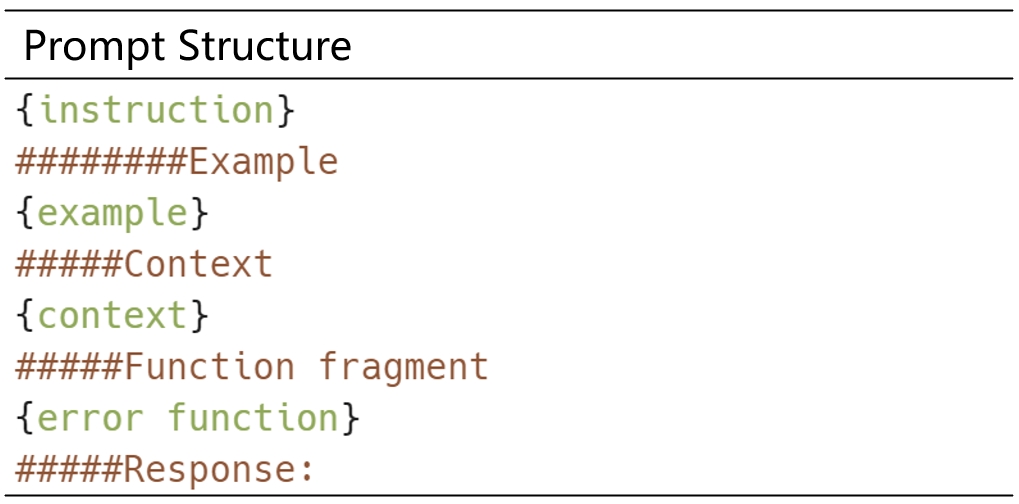}
  \caption{Prompt structure of our experiment}
  \label{img:prompt structure}
\end{figure}

We prepare Figure~\ref{img:prompt structure}, illustrating the design structure of prompts in our experiments. From top to bottom, it includes the task instruction, example (one-shot method), context retrieved by the RLCE method, and the Error Function.

\begin{figure}[h]
  \centering
  \includegraphics[width=.9\linewidth]{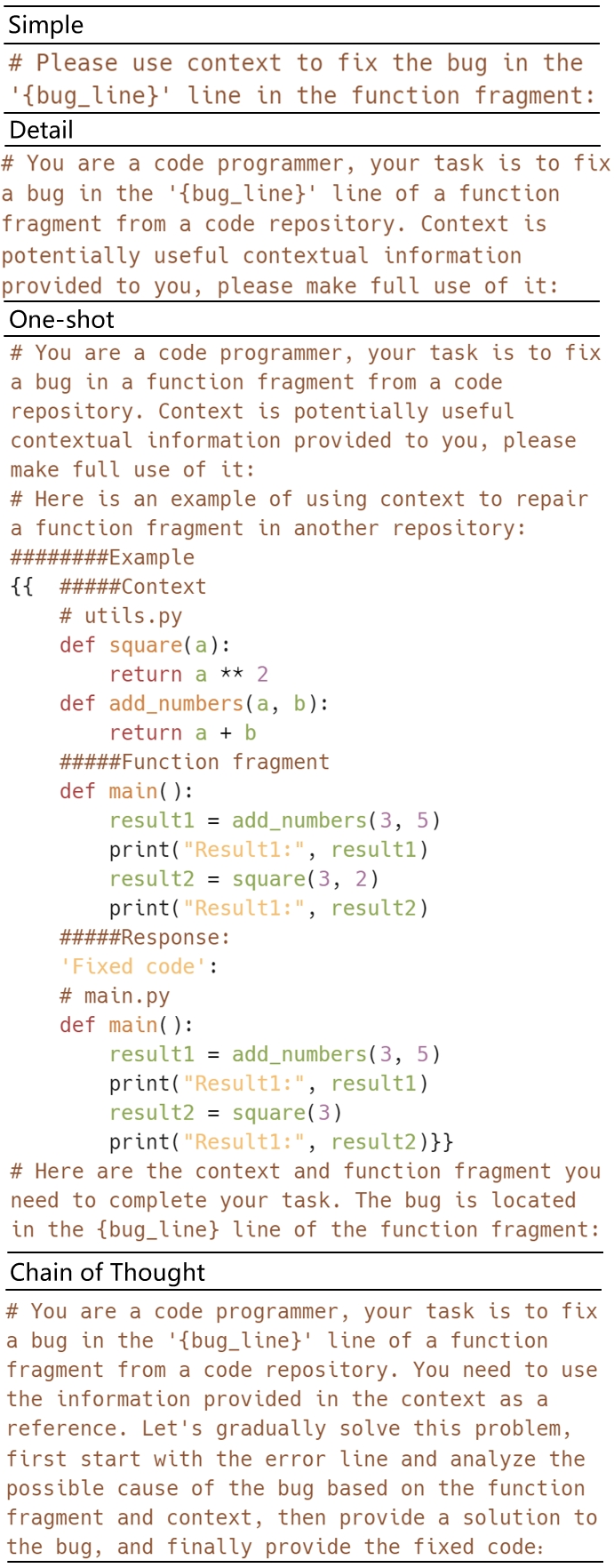}
  \caption{Instructions for different prompt strategies} 
  \label{img:instruction}
  \vspace{14pt}
\end{figure}

For each prompt strategy, the designed instructions are depicted in Figure~\ref{img:instruction}.

\subsection{Case Study}
\subsubsection{Diverse repair methods}
\label{case:one}
\begin{figure}[h]
  \centering
  \includegraphics[width=\linewidth]{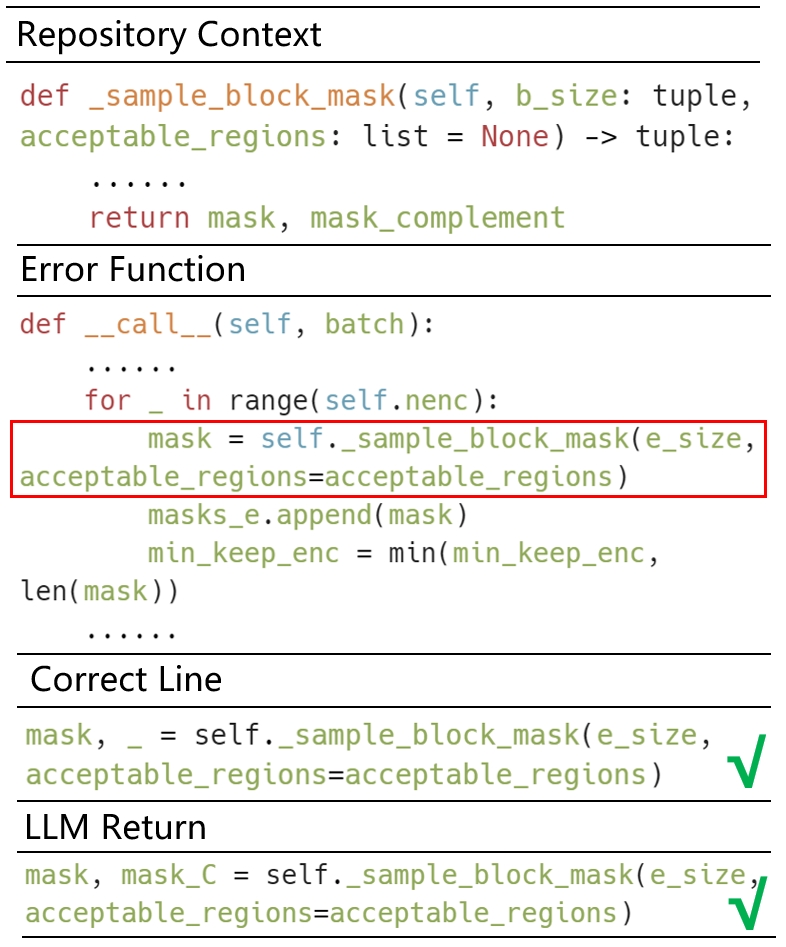}
  \caption{Diversification of repair methods}
  \label{img:case(evaluation)}
\end{figure}

In APR tasks, one bug often corresponds to multiple distinct repair possibilities, increasing the difficulty of assessing the accuracy of model-generated repairs. Figure~\ref{img:case(evaluation)} illustrates an example, the error involves a missing parameter in the parameters returned by the function. The original correct line in the repository utilized ``\_'' to receive this parameter, and both the manual correction and the repair generated by the GPT3.5 model employed ``mask\_C'', effectively repairing the error.

\subsubsection{RLCE for other program languages}
\label{case:two}
\begin{figure}[h]
  \centering
  \includegraphics[width=\linewidth]{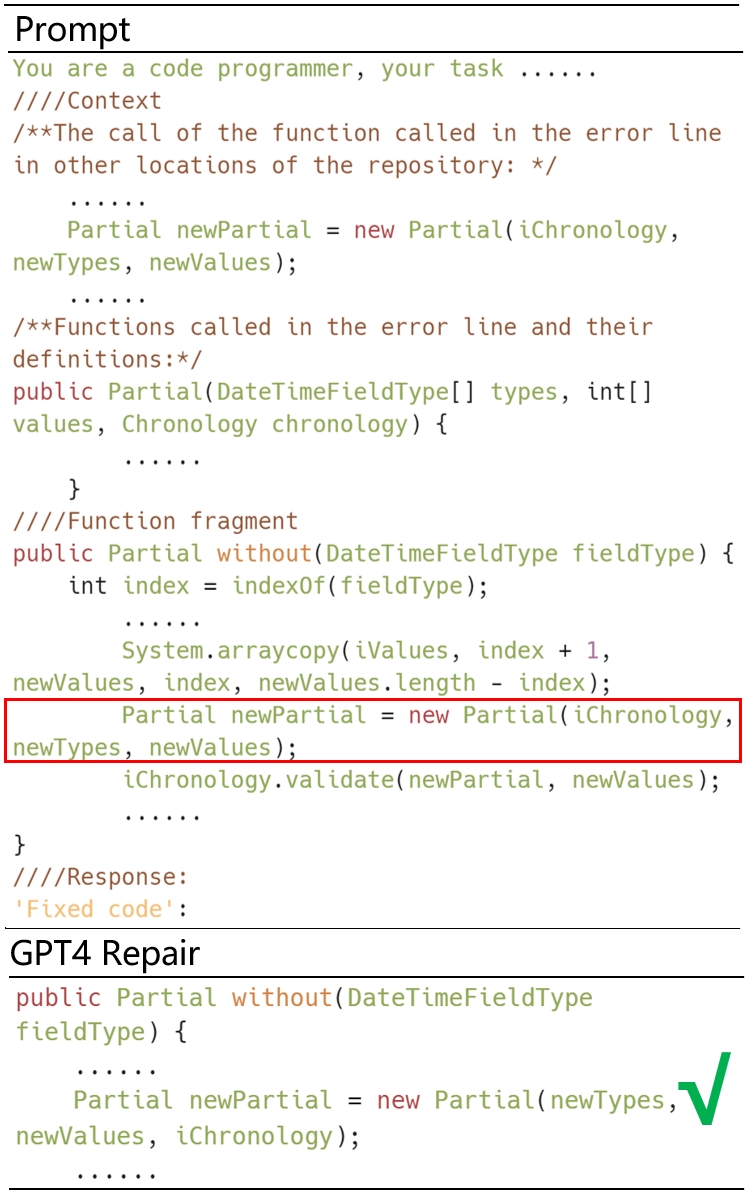}
  \caption{RLCE method for Java language program repair}
  \label{img:case(java)}
\end{figure}

As described in Section 3.3, RLCE is a simple and versatile method that can be applied to different programming languages. As illustrated in Figure~\ref{img:case(java)}, in this example, the bug arises from an incorrect parameter order during a function call. GPT-4 successfully repairs the bug based on the repository-level context provided by our RLCE method.

\clearpage
\bibliographystyle{ACM-Reference-Format}
\bibliography{software}

\end{document}